\documentclass[fleqn,usenatbib]{mnras}

\usepackage{newtxtext,newtxmath}

\usepackage[T1]{fontenc}
\usepackage{ae,aecompl}

\usepackage{graphicx}	
\usepackage{amsmath}	
\usepackage{algorithm2e}
\usepackage[version=4]{mhchem} 
\usepackage{float}
\usepackage{lscape}
\DeclareUnicodeCharacter{2060}{\nolinebreak}

\newcommand{\kms}{km\,s$^{-1}$} 

\title[The impact of shear on cloud rotation]{The impact of shear on the rotation of Galactic plane molecular clouds}

\author[R. Rani et al.]{
Raffaele Rani,$^{1}$\thanks{E-mail: rani@ntnu.edu.tw (RR)}
Jia-Lun Li,$^{2}$
Toby J. T. Moore,$^{3}$
David J. Eden,$^{4}$
Andrew J. Rigby,$^{5}$ 
\newauthor
\  Geumsook Park,$^{6, 7, 8}$ 
Yueh-Ning Lee $^{1}$
\\
$^{1}$Center of Astronomy and Gravitation, Department of Earth Sciences, National Taiwan Normal University, 88, Sec. 4, Ting-Chou Rd., Wenshan District, \\ Taipei 116, Taiwan R.O.C.\\
$^{2}$ Institute of Astronomy, National Tsing Hua University, Hsinchu 30013, Taiwan R.O.C.\\
$^{3}$Astrophysics Research Institute, Liverpool John Moores University, IC2, Liverpool Science Park, 146 Brownlow Hill, Liverpool, L3 5RF, UK\\
$^{4}$Armagh Observatory and Planetarium, College Hill, Armagh, BT61 9DB, UK\\
$^{5}$School of Physics and Astronomy, University of Leeds, Leeds LS2 9JT, UK\\
$^{6}$Telepix Co., Ltd., 17, Techno 4-ro, Yuseong-gu, Daejeon 34013, Republic of Korea\\
$^{7}$Research Institute of Natural Sciences, Chungnam National University, 99 Daehak-ro, Yuseong-gu, Daejeon 34134, Republic of Korea\\
$^{8}$Korea Astronomy and Space Science Institute, 776 Daedeokdae-ro, Yuseong-gu, Daejeon 34055, Republic of Korea
}

\date{Accepted XXX. Received YYY; in original form ZZZ}

\pubyear{2022}

\begin{document}
\label{firstpage}
\pagerange{\pageref{firstpage}--\pageref{lastpage}}
\maketitle

\begin{abstract}

Stars form in the densest regions of molecular clouds, however, there is no universal understanding of the factors that regulate cloud dynamics and their influence on the gas-to-stars conversion. This study considers the impact of Galactic shear on the rotation of giant molecular clouds (GMCs) and its relation to the solenoidal modes of turbulence. We estimate the direction of rotation for a large sample of clouds in the \ce{^{13}CO}/\ce{C^{18}O} (3-2) Heterodyne Inner Milky Way Plane Survey (CHIMPS) and their corresponding sources in a new segmentation of the \ce{^{12}CO}(3-2) High-Resolution Survey (COHRS). To quantify the strength of shear, we introduce a parameter that describes the shear's ability to disrupt growing density perturbations within the cloud.
Although we find no correlation between the direction of cloud rotation, the shear parameter, and the magnitude of the velocity gradient, the solenoidal fraction of the turbulence in the CHIMPS sample is positively correlated with the shear parameter and behaves similarly when plotted over Galactocentric distance. GMCs may thus not be large or long-lived enough to be affected by shear to the point of showing rotational alignment.  In theory, Galactic shear can facilitate the rise of solenoidal turbulence and thus contribute to suppressing star formation. These results also suggest that the rotation of clouds is not strictly related to the overall rotation of the disc, but is more likely to be the imprint of Kelvin-Helmholtz instabilities in the colliding flows that formed the clouds. 

\end{abstract}



\section{Introduction}

The efficiency and rate at which molecular gas is converted into stars determine the evolution and the observable properties of galaxies. Compressive motions associated with large-scale instabilities \cite[gravitational instabilities upon entering regions with differing densities, shear originating from the differential galactic rotation, expanding superbubbles created by supernova explosions][]{Elmegreen1995, McKee2007} induce the transition from atomic to molecular gas in the diffuse interstellar medium ISM and the formation of giant molecular clouds. 

In particular, the Kelvin–Helmholtz Instability (KHI) is a well-known instability arising at the interface of shearing fluids \citep{Drazin1981}. In the ISM, the presence of shearing flows can lead to the formation of KHI and may cause the formation of clouds. KHIs may induce rotation \citep{Fleck1989} and contribute to turbulence and/or change the properties and structures of the molecular cloud through turbulence and mixing \citep{Berne2010, Rollig2011, Sahai2012, Meidt2018, Meidt2019}. 
 These mechanisms lead to the formation of shock-bounded layers between convergent flows, a process that induces
fragmentation through non-linear instabilities \citep{Vishniac1994}. 
Numerical simulations
of this scenario show that fully developed turbulence arises in the shock-driven layers \citep{Hunter1986, Klein1998, Heitsch2009, Inoue2012, Pudritz2013, Inutsuka2015}. 
This turbulent state is maintained throughout the duration of a stream collision and its fragmentation into molecular clouds. 
Cloud structure can also be changed by cloud-cloud collisions, which yield larger and more massive clouds \citep{Tan2000}.

The internal turbulence of molecular clouds originates from a dissipative energy cascade in compressible turbulent flows. At every scale, the fraction of the energy that is not dissipated through shocks is transmitted to smaller-scale structures \citep{Kornreich2000}. In this framework, the relatively high star-formation efficiency (SFE) observed in disc clouds is linked to the prevalence of compressive (curl-free) turbulent modes. In contrast, the
low SFE that characterises clouds in the Central Molecular Zone (CMZ) is related to the shear-driven solenoidal (divergence-free) component \citep{Longmore2013}. A similar analysis of the Orion B molecular cloud \citep{Orkisz2017} found that the turbulence is mostly solenoidal, consistent with the low star-formation rate associated with the cloud. These solenoidal modes are,
however, position-dependent and vary with scale within the Orion B cloud, with motions around
the main star-forming regions being strongly compressive. Thus, significant inter-cloud variability of the compressive/solenoidal mode fractions may be a decisive agent of variations in the SFE. 

This framework suggests that the magnitude of shear in galaxies or the shear-induced cloud-cloud collisions contributes to the SFE and star formation rate \cite[SFR][]{Silk1997, Tan2000}. Models \citep{Tan2000} and simulations \citep{Weidner2010, Colling2018} yield contrasting evidence on the relation between enhanced/reduced SFR and high/low shear. Although, shear-induced cloud-cloud collisions promote higher  SFRs, accounting for magnetic fields and stellar feedback in hydrodynamical simulations seems to complicate the relationship between star formation and shear \citep{Colling2018}. If a weaker strength of shear should enable higher rates of star formation, feedback from the first generation of stars will dissipate the gas in molecular clouds over shorter timescales, resulting in lower SFEs. A study \citep{Watson2012} of 20 bulgeless galaxies showed no correlation between SFE and the galaxies' circular velocities. Irregular galaxies also show poor correlations between the shear magnitude and
the SFRs \citep{Hunter1998}. 

The analysis of shear-quantifying parameters in a sample of clouds in the Galactic Ring Survey \citep{Jackson2006} found no evidence that shear plays a significant role in opposing gravitational collapse. In addition, the shear parameter of the clouds (see sub-section \ref{shear_parameters}) does not depend on the Galactic environment and no correlations between this measure (estimated on a Galactic rotation model) and several indicators of their star formation activity was found \citep{Dib2012}. 

If shear enhances the solenoidal modes of turbulence in clouds by inducing a velocity gradient in molecular clouds, it could be a factor responsible for the observed decline in the relative fraction of power in turbulent solenoidal modes (the solenoidal fraction) with Galactocentric distances \cite[][shear being assumed to be stronger closer to the Galactic centre]{Rani2022} in the\ce{^{13}CO}/\ce{C^{18}O} ($J=3-2$) Heterodyne Inner Milky Way Plane Survey, CHIMPS \cite{Rigby2016}. In this framework, shear could also be a major contributor to the negative correlation between the solenoidal fraction and the SFE in CHIMPS \cite{Rani2022}. If this were the case and shear really induced a full-scale velocity gradient in molecular clouds, we ought to be able to quantify this effect by considering the overall rotation of molecular clouds. Shear-induced rotation would then also introduce full-scale solenoidal modes. However, although a cloud's overall rotation indeed contributes to the increase of the solenoidal modes over the entire structure (see below), local vorticity and the gas motions in larger, looser, more rarefied envelopes also result in higher values of the overall solenoidal modes \citep{Orkisz2017}. The solenoidal fraction, used in our studies to quantify the solenoidal turbulence in fact,  provides an estimate of the relative power in the solenoidal component of turbulence within an individual cloud without distinguishing how these motions originate. 

Any cloud rotation induced by shear related to Galactic rotation would manifest as an alignment of the clouds' axes of rotation with the axis of the Galaxy. This alignment would then be lost with the diminishing impact of shear at larger Galactocentric distances. 


In this work, we consider the sample of CHIMPS clouds produced by \citep{Rani2022} together with a novel sample of \ce{^{12}CO}(3-2) sources in the CO High Resolution Survey, COHRS \citep{Dempsey2013, Park2022} to quantify the strength of shear through induced cloud rotation. Furthermore, making use of shear parameters defined by \cite{Dib2012}, we attempt to find a connection between shear, SFE and the solenoidal fraction in CHIMPS clouds. 

Section \ref{data} provides a brief description of the surveys chosen and the construction of a new COHRS catalogue, including emission extraction and distance assignments. Section \ref{methods}
introduces the methods used to identify velocity gradients and the directions of the rotation axes in first-moment maps. It also provides the definition of a parameter used to quantify the effect of shear. Finally, the results of the analysis are presented and discussed in Sections \ref{results} and \ref{conclusions}.

\section{Data}
\label{data} 

\subsection{Surveys}\label{surveys}

The \ce{^{13}CO}/\ce{C^{18}O} ($J=3-2$) Heterodyne Inner Milky Way Plane Survey (CHIMPS) is a spectral survey of the  $J = 3 - 2$  rotational transitions of \ce{^{13}CO} at 330.587\,GHz and \ce{C^{18}O} at 329.331\,GHz.
The survey encompasses approximately 19 square degrees of the Galactic plane, ranging in longitudes from $27\fdg 5$ to $46\fdg 4$ and with latitudes $|b| < 0\fdg5$. It was conducted with an angular resolution of 15 arcseconds using observations carried out at the James Clerk Maxwell Telescope (JCMT) in Hawaii \citep{Rigby2016}. Both 
isotopologues were observed concurrently \citep{Buckle2009} using the  Heterodyne Array Receiver Programme (HARP) together with the Auto-Correlation Spectral Imaging System (ACSIS). The data collected are structured into position-position-velocity (PPV) cubes, each with velocity channels of 0.5\,km\,s$^{-1}$ and a bandwidth of 200\,km\,s$^{-1}$. The velocity range varies, covering from $-50 < v_{\rm LSR} < 150$ km\,s$^{-1}$ at $28^\circ$ to $-75 < v_{\rm LSR} < 125$ km\,s$^{-1}$ at $46^\circ$, in order to accommodate the Galactic velocity gradient associated with the spiral arms in the kinematic local standard of rest \citep[e.g.,][]{Dame2001}. The mean root mean square  (rms) sensitivities of the \ce{^{13}CO} survey are $\sigma(T_{\rm A}^{*})\approx 0.6$\,K per velocity channel, while for \ce{C^{18}O} we have $\sigma(T_A^{*})\approx 0.7$\,K, where $T_A^{*}$ represents the antenna temperature corrected for ohmic losses inside the instrument, spillover, rearward scattering, and atmospheric attenuation \citep{Rigby2016}.

The CO Hi-Resolution Survey (COHRS) mapped the \ce{^{12}CO} ($3-2$) emission in the Inner Milky Way plane, covering latitudes $10 \fdg 25 < l < 17\fdg 5$ with longitudes $|\,b\,| \leq 0\fdg 25$ and $17\fdg 5 < l < 50\fdg 25$  with $|\,b\,| \leq 0\fdg 25$ \citep{Park2022}.  
This particular region was selected to match a set of
important surveys including CHIMPS, the Galactic
Ring Survey \cite[GRS;][]{GRS}, the FOREST Unbiased
Galactic plane Imaging survey with the Nobeyama 45-m
telescope  \cite[FUGIN; ][]{FUGINI}, the Galactic Legacy
Infrared Mid Plane Survey Extraordinaire  
\cite[GLIMPSE;][]{GLIMPSE}, the Bolocam Galactic Plane Survey \cite[BGPS; ][]{bolocam}, and the {\em Herschel} Infrared Galactic Plane Survey \cite[Hi-GAL; ][]{higal}. COHRS observations were also performed at JCMT with HARP at $345.786$\, GHz and ACSIS set to a 1-GHz bandwidth, yielding a frequency resolution of 
$0.488$\,MHz ($0.42$\,km\,s$^{-1}$). The survey covers a velocity
range between  $-30$ and $155$\,\kms, with a spectral resolution of
$0.635$\,km\,s$^{-1}$ and angular resolution of $16.6$\, arcsec
(FWHM), producing a mean rms of $\approx 0.7$\,K in $T_A^{*}$.
 
The COHRS data (second release) are publicly available\footnote{\url{https://doi.org/10.11570/22.0078}}. In our analysis, we construct a sub-sample of the full set of COHRS sources by only considering those COHRS sources that contain the emission peaks of CHIMPS clouds. This set comprises 452 clouds. Hence this catalogue will be simply referred to as the COHRS catalogue/survey.

\section{Constructing a new COHRS catalogue}

\subsection{Cloud extraction}\label{cloud_extraction}

To identify molecular clouds in the COHRS data, we employ
the Spectral Clustering for Interstellar Molecular Emission Segmentation (SCIMES) algorithm \citep{Colombo2015}. This image-segmentation method encodes the global hierarchical structure of emission within a molecular-line datacube into a dendrogram.
This emission dendrogram is produced through the  Python package for astronomical dendrograms \cite[Astrodendro, ][]{astropy:2013, astropy:2018}. SCIMES then uses similarity criteria to analyse the dendrogram by
recasting it as a weighted complete graph (with vertices
corresponding to the leaves of the dendrogram and weighted edges
representing the chosen affinity relations between the leaves).
The SCIMES algorithm then uses spectral clustering on the Laplacian of the affinity matrix representing the graph to partition the dendrogram into separate components. These clusters define a segmentation of the emission into individual clouds.  

Because of the variable weather conditions and the varying number of active receptors during the 4 semesters of COHRS observations, the original cubes do not present an entirely uniform
sensitivity across the whole survey \citep{Park2022}. To avoid high-noise regions being incorrectly identified as
clouds (i.e. false positives) and to prevent the loss
of real signal-to-noise sources in regions of low background (false negatives), the SCIMES extraction is performed on the signal-to-noise ratio
(SNR) cubes instead of brightness-temperature data
\citep{Moore2015, Eden2017}.  This novel COHRS catalogue thus differs from the one constructed by \cite{Colombo2018} for the first COHRS data release for which the actual emission maps (masked for given SNR thresholds) were used for the SCIMES segmentation.

Since the area covered by both CHIMPS and COHRS is too
large to be analysed as a single datacube and since our analysis of
shear focuses on the largest \ce{^{12}CO} structures in the emission as they are
more likely to be impacted by the differential rotation of the
Galaxy, we organise the original datacubes into 18 regions of $4^{\circ} $ longitude each, with resolution degraded by a factor 2 in all 3 axes. 
Each pair of adjacent regions overlaps by $2^{\circ}$. This wide
overlap allows for any source in contact with the edges of
any region to be completely included
and accounted for in the adjacent one (see below).  
After the extraction, only clouds that contained at least one voxel with SNR $\geq 10$ are retained.
The parameters defining the emission dendrogram are
chosen as multiples of the background 
$\sigma_\mathrm{rms}$ (with $\sigma_\mathrm{rms} = 1$ in
SNR cubes). We set each branch of the dendrogram to be defined by a SNR change (\texttt{min\_delta⁠}) of $3$ and to
contain at least 5 voxels ($\mathtt{min \_npix}=5$⁠). Any value of the SNR below $3\sigma_{\mathrm{rms}}$ (⁠$\mathtt{min \_val} = 3 $⁠) is not considered.

\begin{figure*}
	\includegraphics[width=0.9\textwidth]{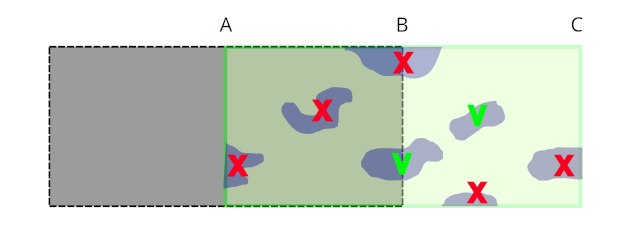}	
	\caption{Prescription for cloud removal in the overlapping area of adjacent regions (between edge A and B in the figure). In each region, we remove all clouds in contact with the edges of the field of observation. Clouds that touch the B edge are assigned to the right region (light green area), while those t between edges B and C are assigned to the next region (grey and shaded green areas). The process is repeated on all regions with the exception of the last region on the right, for which all the clouds between A and B are kept. Green ticks and red crosses indicate the clouds that are removed and those that are left, respectively.}
	\label{rcuts} 
\end{figure*}

The catalogue is rid of spurious sources and noise artefacts that are left after extraction by applying an additional filter. This mask removes those sources that either cover fewer than 4 pixels in each spatial direction or four velocity channels. 
This requirement ensures that each cloud is fully resolved in each direction and that the selection does not include sources with too small a field size.

Clouds touching the edges of the field of observation are removed from the catalogues. To avoid double counting of sources in the overlapping areas between regions, the sources to remove in each region are selected according to the recipe shown in Figure \ref{rcuts}. The two-degree overlap was chosen after visual inspection to ensure that the recipe would yield a unique set of sources for all the 18 regions COHRS is divided into. The final COHRS catalogue comprises 3271 sources.

\subsection{Distances}\label{distances}

The distances and physical properties of the sources identified through the segmentation of CHIMPS are taken from the SCIMES catalogue introduced in \cite{Rani2023}. 
Helio- and Galactocentric distances to COHRS sources are assigned through direct comparison with the aforementioned CHIMPS catalogue. COHRS and CHIMPS sources are matched by considering the position of the COHRS peak emission in the CHIMPS segmentation assignment cubes. A source whose peak lies within a CHIMPS cloud is ascribed the distance of that cloud.  Unassigned sources are not required in the present study, however they can be givendistances by calculating their position with a Galactic rotation curve model \citep{Brand1993, Reid2016}. 

\subsection{Size and mass}\label{sizes}

We estimate the size of the CHIMPS and COHRS sources by considering an `equivalent' radius that approximates the spatial extension of a cloud. Following \cite{Rigby2019}, we define the equivalent radius $R_\mathrm{eq}$  as the radius of the circle whose area ($A_c$) is equivalent to the projected area of the source scaled to its distance, 

\begin{equation}\label{req}
    R_\mathrm{eq} = \sqrt{A_c/\pi}.
\end{equation}

\begin{figure}
	\includegraphics[width=1.1\columnwidth]{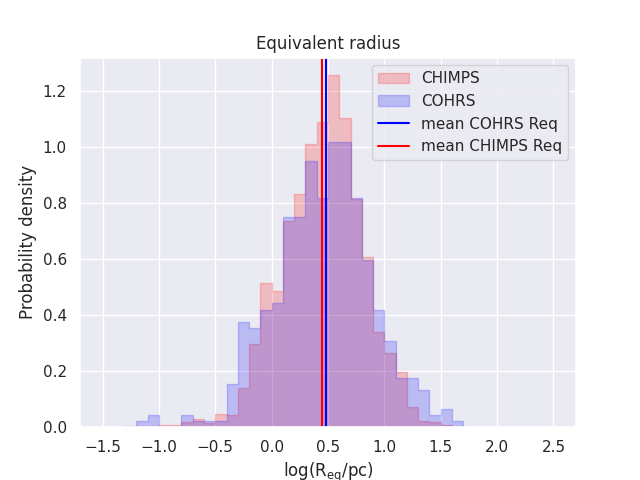}	
	\caption{Distribution of equivalent radii across CHIMPS (red) and COHRS (blue) matching samples. The vertical lines indicate the mean values.
 }
	\label{histo_req} 
\end{figure}

The values of the equivalent radii associated with the SCIMES and COHRS sources were calculated directly from the values of the exact areas produced by the {\sc Astrodendro} dendrogram statistics tools. 
Figure \ref{histo_req} displays the distribution of the equivalent radii of sources in the two surveys. Although $^{12}${CO}\,(3--2) emission detected some slightly larger structures than $^{13}${CO}\,(3--2), the mean size of the sources in the two surveys takes on similar values. We notice that the lower optical depths of \ce{^{13}CO} imply that, although these clouds are more centrally concentrated, $R_\mathrm{eq}$ does not account for the intensity distribution, and so it returns an upper limit for the cloud size. 
Due to the high opacity of \ce{^{12}CO} and a resulting lower effective critical density, the \ce{^{12}CO} transition probes significantly lower \ce{H_2} volume densities in the clouds than \ce{^{13}CO} \citep{Tang2013, Roueff2020, Mazumdar2021}. Therefore, the COHRS map delineates broader scales, whereas the emission of \ce{^{13}CO} is responsive to gas with higher column density, thus delineating the denser, more compact clumps.
As larger structures are expected to be more susceptible to the impact of shear, the COHRS sample is instrumental in revealing any preferential rotation induced by shear.

The mass of CHIMPS clouds can be then estimated through the column-density cubes \citep{Rigby2019}. The \ce{H_2} mass of the cloud is estimated by considering the mean mass per \ce{H_2} molecule, taken to be 2.72 times the mass of the proton, accounting for a helium fraction of 0.25 \citep{Allen1973}, and an abundance of $10^6$ \ce{H_2} molecules per \ce{^{13}CO} molecule \citep{Draine2011}.

COHRS masses are expressed in terms of the molecular gas luminosity and calculated through a \ce{^12}{CO} luminosity-mass conversion factor $M = \alpha_\mathrm{CO} L_\mathrm{CO}$, with
$\alpha_\mathrm{^{12}CO(1-0)} = 4.35\,\mathrm{  M}_\odot\, \mathrm{    pc}^{-2}\,\mathrm{km}^{-1}\mathrm{s}$, assuming a mean molecular weight of $2.72 m_\mathrm{H}$ per hydrogen molecule \citep{Colombo2018}. 


\begin{figure}
	\includegraphics[width=1.1\columnwidth]{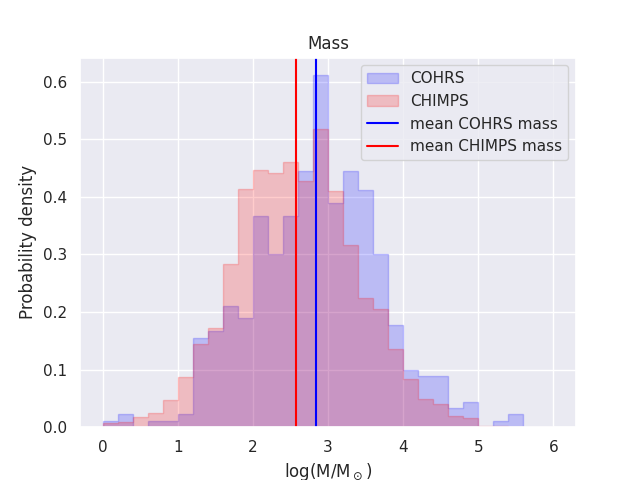}	
	\caption{Distribution of masses across CHIMPS (red) and COHRS (blue). The vertical lines indicate the mean values.}
	\label{histo_mass} 
\end{figure}

The distribution of masses in COHRS (Fig. \ref{histo_mass}) peaks at higher values and exhibits an overall greater number of sources at higher mass values than CHIMPS. This behaviour may be enhanced by the different methods used in the mass estimation. 

\section{Cloud rotation, shear and turbulence}

\subsection{Velocity gradients and rotation angles}\label{methods}

\begin{figure*}
	\includegraphics[width=0.65\textwidth]{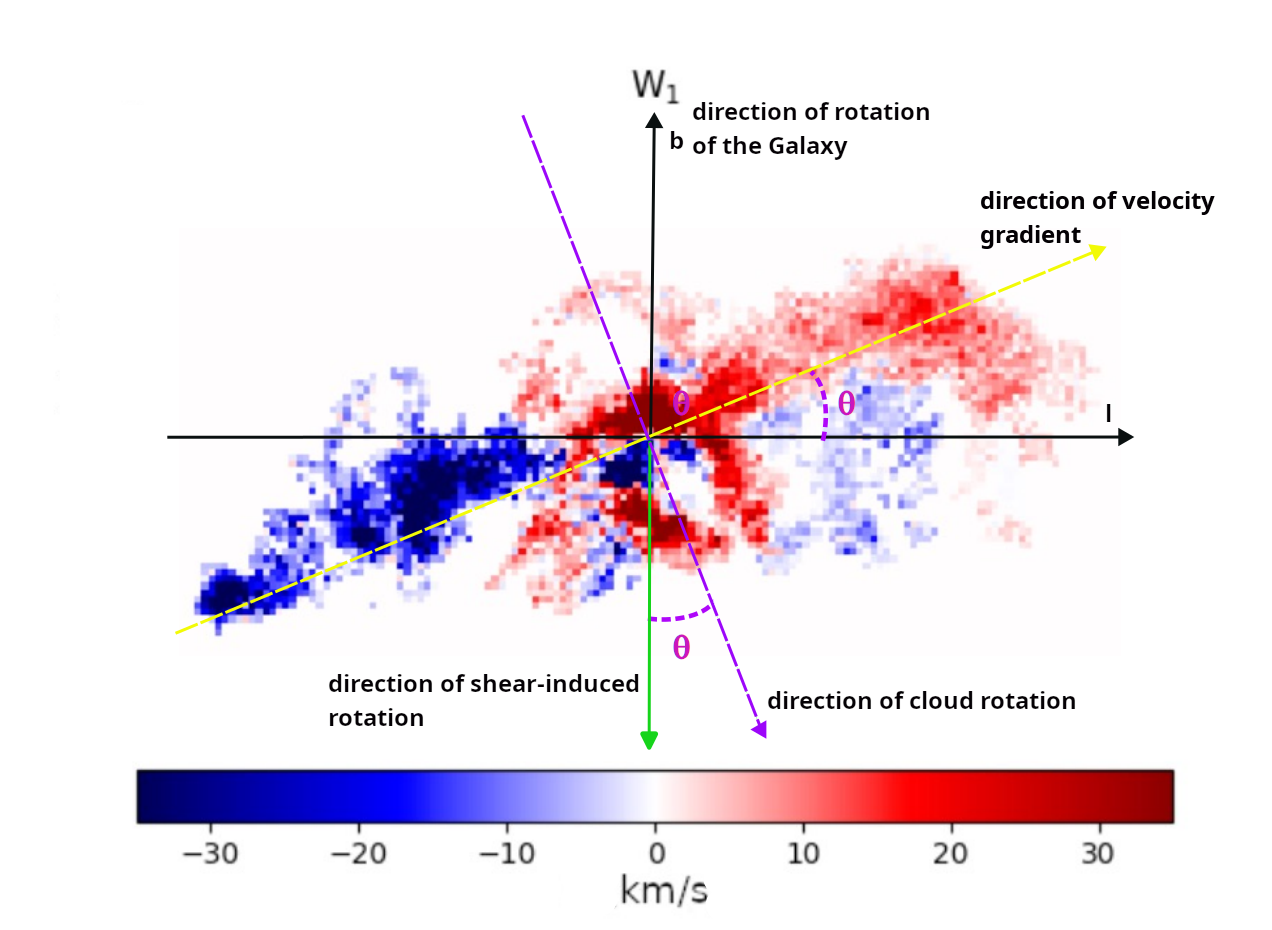}	
	\caption{Definition of the direction of rotation of a cloud (see text). The angle $\theta$ is defined with reference to the direction of rotation induced by Galactic shear.}
	\label{skx} 
\end{figure*}

If shear is the dominant force behind cloud rotation, at least at short 
Galactocentric distances where shear is stronger, we should expect clouds to rotate with axes parallel to the axis of rotation of the Galaxy (but with opposite direction, see Fig. \ref{skx}).
Cloud rotation gives rise to a velocity gradient across a molecular cloud. Although several processes \cite[such as turbulence][]{Buckert2000} may induce velocity gradients on different scales in molecular clouds, we assume a full-scale gradient be an indicator of cloud rotation as it may be induced by Galactic shear. Measuring the overall velocity gradient is the most reliable and systematic approach to the detection of rotation in large samples of clouds \citep{Braine2018}. 
The evaluation of the velocity gradient and the subsequent
determination of the direction of the cloud’s axis of rotation is
performed in two steps. First, for each cloud, we calculate the first
moment of the velocity of the emission map in the frame of reference
of the centre of mass of the cloud (thus by shifting the velocity axis
by the velocity of the cloud’s centroid)

\begin{equation}
    v(l,b) = \frac{\sum_i I_i v_i dv}{\sum_i I_i dv}, 
\end{equation}

\noindent
where all non-zero emission voxels $i$ in the cloud contribute to the summations with their velocity $v_i$ and emission $I_i$. All velocity channels have the same size $dv$.
This yields an emission-weighted velocity value for each position ($l,b$) in the cloud, 
Considering a frame of reference centred at the average values of the map, we now fit a plane \citep{Imara2011}:

\begin{equation}\label{plane}
  v(l,b) = C_1 l + C_2b + C_3
\end{equation}

\noindent
with $C_1= \partial v / \partial l$, $C_2= \partial v / \partial b$, being the velocity gradients in $l$ and $b$, and $C_3$ is a constant ($C_3 = 0$ in the frame of reference of the centre of mass of the cloud).

The magnitude of the linear velocity gradient is then

\begin{equation} \label{magnitude}
    \Omega = \frac{(C_1^2 + C_2^2)^{\frac{1}{2}}}{d},
\end{equation}

\noindent
where $d$ is the distance of the source (see also Fig. \ref{sol} and \ref{fig9}).

The direction of the velocity gradient (direction of increasing velocities) with respect to the positive direction of  Galactic longitude axis (in the $l-b$ plane) can be estimated as the angle

\begin{equation}\label{rotation}
    \theta=  \arctan\left(\frac{C_2}{C_1}\right).
\end{equation}

The angle $\theta$, measured in degrees also represents the angle between the axes of cloud rotation, perpendicular to  the direction of the velocity gradient, and the rotation of the Galaxy as shown in Fig. \ref{skx}. 

Since we lack information on the inclination $i$ of the individual clouds to our line of sight, the measurement of the velocity gradient underestimates the actual value by a factor $\sin(i)^{-1}$ \citep{Phillips1999, Imara2011}. Thus, the values of $\theta$ calculated through equation \ref{rotation} represent a lower bound for the sample, with the alignment with the Galactic rotation axis worsening when the actual inclination of the cloud rotation is considered. 
The lack of information on cloud inclination on the other hand leads to underestimating the measured magnitude of the velocity gradient ($\Omega$) and, consequently, the angular velocity of a cloud's rotation \citep{Imara2011}, making these measures much less reliable than the rotation angle for the proposed analysis of shear . The method presented in this section focuses on determining the direction of rotation of each cloud in the sample. As the direction of the axis of rotation is ultimately determined by the overall direction of the velocity gradient, this, in turn, accounts for the strengths of the internal motion of the substructure within the cloud.

In our study, we consider the angle  $\theta$. defined in Fig.\,\ref{skx} as $\theta = 90^\circ - \tilde{\theta}$, with $\tilde{\theta}$ derived from fitting the first-moment map with a plane, as in equations \ref{plane} and \ref{rotation}.

The angle $\theta$ quantifies the difference in orientation between the direction of cloud rotation and the direction of shear-induced rotation. 

\subsection{Quantifying the impact of shear}\label{shear_parameters}

Assuming that overdensities in the molecular ISM accrete mass through streaming motions along magnetic field lines, their growth then depends solely on the interplay between gravity and the local level of galactic shear \citep{Elmegreen1993,  Hunter1998}. In this framework, since magnetic
fields can transfer the angular momentum away from the cloud, self-gravity and shear play a more  relevant role in core formation than the competition between self-gravity, pressure, and the Coriolis force (which is represented
by the Toomre-Q parameter \citep{Toomre1964}. 

In a scenario in which mass accretion i is determined by the competition between their self-gravity and the local strength of Galactic shear, the growth rate of the density perturbations can be expressed as   

\begin{equation}\label{growth_rate}
  P_r =  \frac{\pi G \Sigma}{\sigma_v} \ {\rm s}^{-1},
\end{equation}

\noindent
where $\Sigma$ is the local gas surface density, $\sigma_v$ the velocity dispersion, and $G$ the gravitational constant. Perturbations grow most efficiently when $P_r < |1/A|$,  where $A$ is the Oort constant

\begin{equation}\label{oort}
    A = 0.5\left(\frac{V}{R_\mathrm{gc}} - \frac{dV}{dR_\mathrm{gc}} \right), 
\end{equation}

\noindent
where $R_\mathrm{gc}$ is the Galactocentric distance of the source and $V$ the rotation velocity of the gas at a given Galactocentric radius.  In particular, for a source of size $L$ with centroid located at $R_\mathrm{gc}$.

\begin{equation}\label{cloud_oort}
    A = 0.5 \left( \frac{V(R_\mathrm{gc})}{R_\mathrm{gc}} - \frac{|V(R_\mathrm{gc}+L/2)- V(R_\mathrm{gc}-L/2)|}{L} \right).
\end{equation}

From the Oort constant, we can define a critical surface density; this is the threshold beyond which a density perturbation in the ISM becomes so significant that it cannot be erased by shear:

\begin{equation}\label{critical_surface_density}
 \Sigma_\mathrm{sh} = \frac{\alpha_A A \sigma_v}{\pi G}, 
\end{equation}

\noindent
with $\alpha_A = \ln(Q)/2$. The factor $Q$ represents the growth factor that perturbations in the diffuse ISM must have to overcome the disruptive impact of shear. \cite{Hunter1998} quantified $Q$ to equal 100. This value yields the density contrast between the diffuse ISM with densities $\sim 0.1 - 1\,\mathrm{cm}^{-3}$ and the molecular phase $\gtrsim 100\,\mathrm{ cm}^{-3}$. 

The critical surface density can thus be used to define a shear parameter for gravitational instabilities:

\begin{equation}\label{shear_parameter}
    S = \frac{\Sigma_\mathrm{sh}}{\Sigma} = \frac{\alpha_A A \sigma_v}{\pi G \Sigma}, 
\end{equation}

Shear will disrupt density perturbations at values of $S > 1$ while being ineffective at $S  < 1$.

The quantities that enter the  definition of $S$ above are taken from the catalogues produced by SCIMES (in particular the velocity dispersion and the exact area of each source, adjusted for the distance). 
The masses in CHIMPS are derived from the column density data cubes by scaling each voxel in the source with the assigned distance. The surface density is then taken to be the ratio between the mass and the exact projected area of the source \citep{Rani2023}.   COHRS surface densities are calculated using the luminosity masses reported in the COHRS catalogue \citep{Park2022}.

\subsection{Solenoidal fraction}\label{R}

Our turbulence analysis is based on the 
the statistical method developed by \cite{Brunt2010} and \cite{Brunt2014}, which allows us to quantify the relative fraction of power in the solenoidal modes of turbulence present in a molecular cloud from emission and column density observations. The main idea behind the method is to reconstruct the properties of a three-dimensional source from the information contained in its observed two-dimensional line-of-sight projection. 

According to the Helmholtz theorem, a three-dimensional vector field can be decomposed into a divergence-free (solenoidal or transverse).
component and a curl-free (compressive, parallel) component.  It can be shown through
symmetry arguments and the local orthogonality of the Helmholtz components that, under the assumption of statistical isotropy, the projected line-of-sight component of the three-dimensional field is proportional to the solenoidal component of the full field in $k_z=0$ cut in Fourier space. The solenoidal fraction or the amount of power in the solenoidal modes of the full field is then defined as the ratio of the variances of the projected component and the full field, 
which can be written in terms of the power spectra of these fields (2D and 3D) through the Parceval theorem. Therefore,  the power spectrum of the projected component is a direct measure of the power spectrum of the solenoidal component of the full field. 
In the case of optically thin gas, this relation translates naturally into a function of the power spectra of the zeroth ($W_0$) and first ($W_1$) moments of velocity in which the three-dimensional field in question is the momentum density',$\rho \mathbf{v}$.  
In emission data cubes, the solenoidal fraction, $R$ of each molecular cloud can be practically estimated as

{
\begin{equation}\label{b3}
R = \Bigg[\frac{\langle W_1^2\rangle}{ \langle
W_0^2\rangle}\Bigg]\Bigg[\frac{\langle W_0^2 / \langle W_0
\rangle^2 \rangle}{1 + B_1(\langle W_0^2\rangle/ \langle W_0 \rangle^2 -1)} \Bigg]
\Bigg[ g_{21} \frac{\langle W_2 \rangle}{\langle W_0 \rangle}
\Bigg]^{-1}B_2
\end{equation}

\noindent
where 

\begin{equation}\label{b4}
     B_1 = \frac{(\sum_{k_x}\sum_{k_y}\sum_{k_z}f(k))-f(0)}{\sum_{k_x}\sum_{k_y}f(k))-f(0)},
\end{equation}

and

\begin{equation}\label{b5}
     B_2 = \frac{\sum_{k_x}\sum_{k_y}\sum_{k_z}f_\perp(k)\frac{k_x^2 + k_y^2}{k^2}}{\sum_{k_x}\sum_{k_y}f _\perp (k)}.  
\end{equation}

\begin{figure*}
	\includegraphics[width=1.0\textwidth]{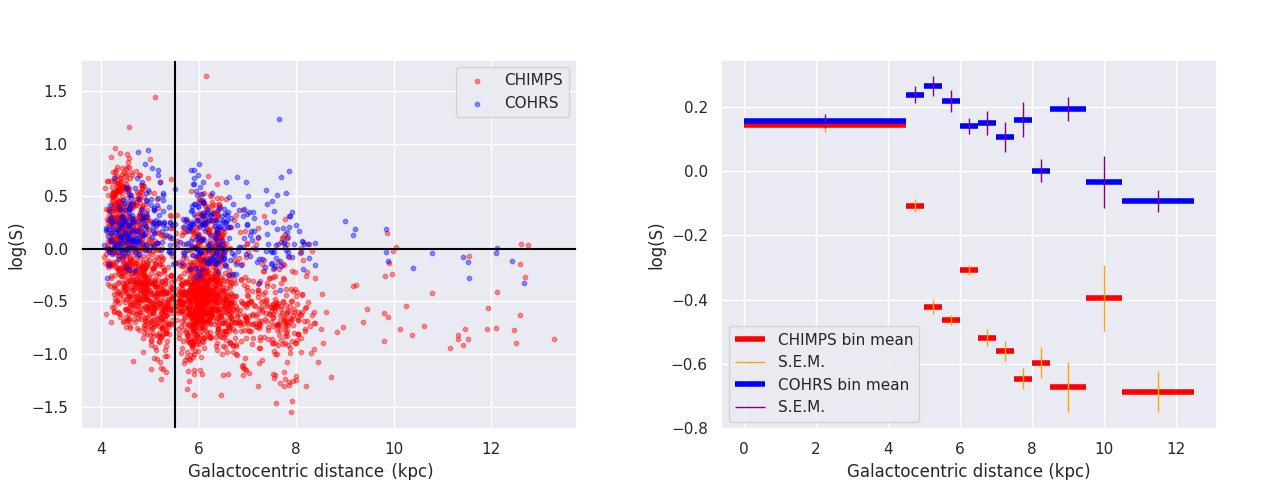}	
	\caption{Distribution of the shear parameter $S$ with Galactocentric distance in the CHIMPS (red) and COHRS (blue) samples. The vertical and horizontal solid black lines mark the $5.5$-kpc and $S =1$ values respectively. These reference values are used to define a $R_\mathrm{gc}$-limited and an $S$-limited subsample of CHIMPS sources (see text). The right panel shows the binned means and the standard errors of the mean (S.E.M.) of the distributions of $S$ over the Galactocentric distance.}
	\label{S_master} 
\end{figure*}

\noindent
with $f(k)$ and $f_\perp(k)$ being the angular (azimuthal) averages of the power spectra of the zeroth and first moments \cite[notation after ][]{Orkisz2017}.
The angle brackets denote the spatial averages of the velocity moments $W_0$, $W_1$, and $W_2$, calculated in the reference frame of the centre of mass of the cloud. 
The statistical correction factor $g_{21}$  accounts for
the correlations between the variations of $\rho$ and
$\mathbf{v}$. If  $\rho$ and $\mathbf{v}$ are uncorrelated,
$g_{21} = 1$. The constant $g_{21}$ can be expressed  In terms of density, velocity and the spatial average of
the density $\rho_0$, by the variance of
the three-dimensional volume density $\langle (\rho /
\rho_0)^2 \rangle$

\begin{equation}\label{b6}
    g_{21} = \frac{\langle\rho^2v^2\rangle/\langle 
    \rho^2\rangle}{\langle \rho v^2\rangle/\langle \rho
    \rangle} =  \Bigg \langle \frac{\rho^2}{\rho_0^2} \Bigg
    \rangle^\epsilon, 
\end{equation}

where $\epsilon$ is a 
is a small positive constant which is the exponent of the power law expressing the relation between the variance of the velocity $\sigma_v^2$ and the density $\rho$ \citep{Brunt2010, Brunt2014}.
In our study of the CHIMPS sample, we use the values of the solenoidal fraction published by \cite{Rani2022}. As the COHRS \ce{^{12}CO} emission violates the requirement of being optically thin, the COHRS source will not be considered when discussing the solenoidal fraction.

\section{Results}\label{results}

Fig.\,\ref{S_master} shows how the shear parameter $S$ changes with Galactocentric distance in the CHIMPS and COHRS samples. The plot suggests that in both samples shear has a stronger impact on clouds at shorter distances from the Galactic centre, declining outwards. 
 Fig.\,\ref{S_master} also emphasises source-crowding within the main spiral arms, with large peaks in source concentration at $\sim 4.5$ and $\sim 6.5$\,kpc. These are the locations of the Scutum and Sagittarius arms seen from the Galactic Centre. The smaller peak at $\sim 7.5$\,kpc corresponds to the Perseus arm. These areas encompass the broadest ranges of $S$. The spikes in $S$ observed at spiral arms location result from the larger variety of clouds collected in these regions.
 

\begin{figure*}
	\includegraphics[width=1.0\textwidth]{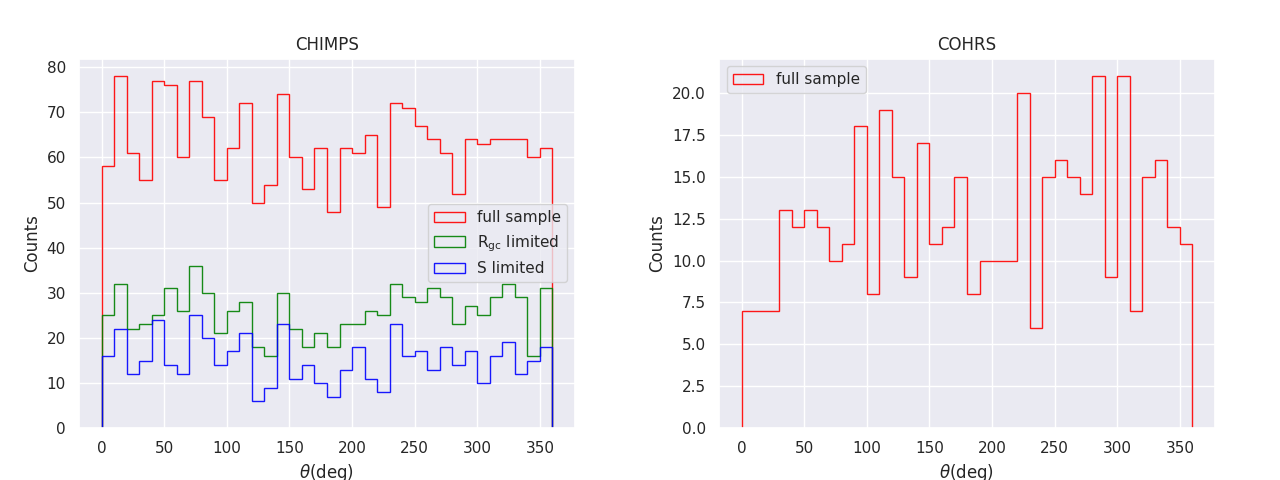}	
	\caption{Distributions of the inclinations of the axes of rotation of the sources in the CHIMPS (left panel) and COHRS (right panel) samples with respect to the axis of rotation of the Galaxy.}
	\label{histo_theta} 
\end{figure*}

The greatest change in $S$ occurs between $3.5$ (the shortest Galactocentric distance probed by CHIMPS) and $5.5$\,kpc. COHRS sources exhibit larger
values of $S$ overall, but exhibit the same trend with distance as their CHIMPS counterparts. In both samples, $S$ peaks at the inner edge of the Inner Galaxy ($\sim 3$\,kpc) and falls off with increasing distance from the Galactic centre. This trend seem to reflect the behaviour of the solenoidal fraction described by 
\cite{Rani2022}, where the relative fraction of power in the solenoidal modes of turbulence peaks at $3.5-4$ kpc, the edge of the region swept by the rotation of the Galactic bar, and declines outwards with a shallow gradient (see Appendix \ref{R_x_chimps}). The similarity in the behaviour of $R$ and $S$ with Galactocentric distance suggests a connection between them.

\begin{figure*}
	\includegraphics[width=1.0\textwidth]{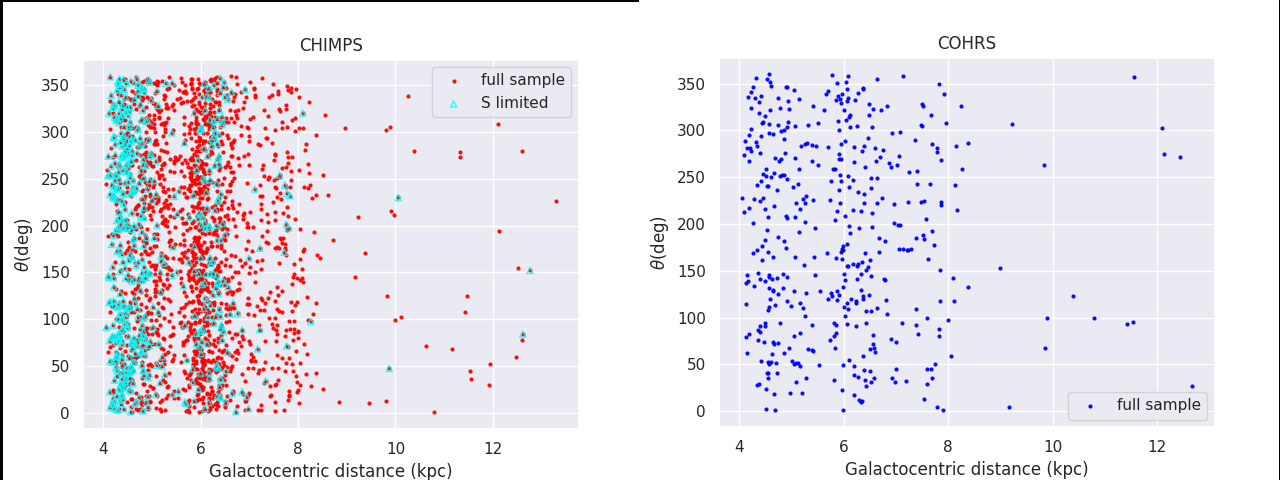}	
	\caption{Distributions of rotation angles with Galactocentric distance.}
	\label{theta_galcen} 
\end{figure*}

\begin{figure}
	\includegraphics[width=1\columnwidth]{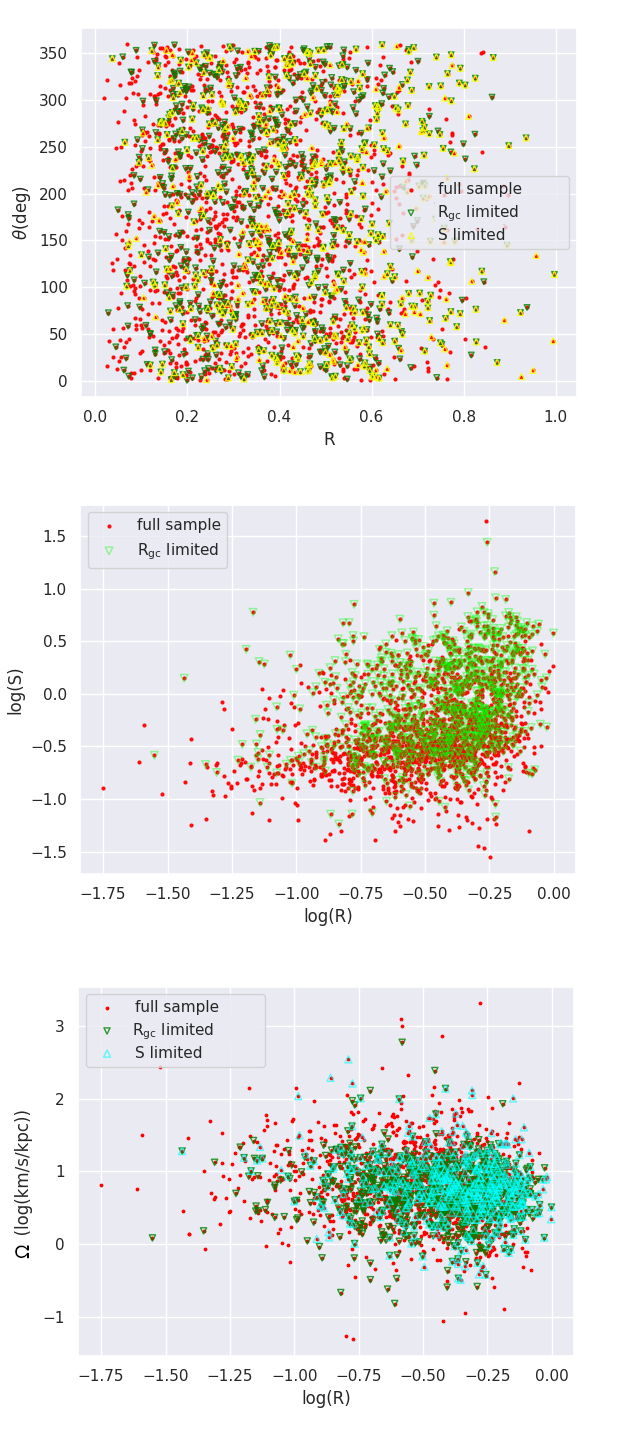}	
	\caption{Top panel: 
 Top panel: Cloud rotation angles (as defined in Fig. \ref{skx} as a function of the solenoidal fraction $R$ for the full CHIMPS survey (red dots) and $R_\mathrm{gc}$-limited subsample (lime triangles).
 Middle panel: The positive correlation between the shear parameter ($S$) and the solenoidal fraction ($R$) in CHIMPS clouds. The picture highlights the higher values of both $S$ and $R$ in the $R_\mathrm{gc}$-limited sample.
 Bottom panel: Distribution of the magnitude of the velocity gradients as a function of the solenoidal fraction $R$ for the full CHIMPS survey (red dots), the $R_\mathrm{gc}$-limited subsample (green triangles), and the S-limited samples (cyan triangles).}
	\label{sol} 
\end{figure}

\begin{figure*}
	\includegraphics[width=1.0\textwidth]{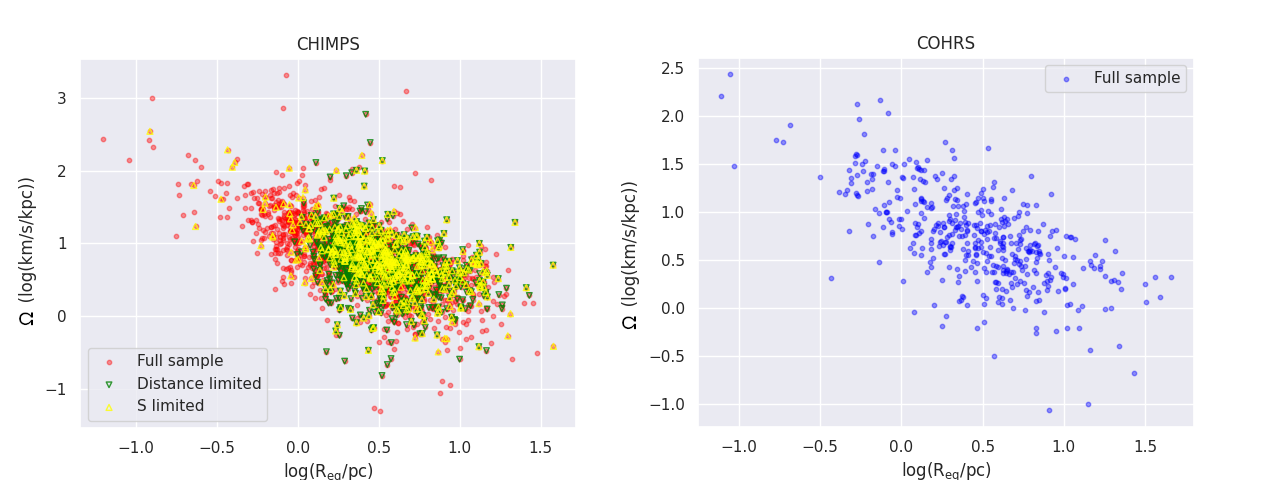}	
	\caption{Left panel: Distribution of the magnitude of the velocity gradients as a function of the equivalent radii of the sources for the full CHIMPS survey (red dots), the $R_\mathrm{gc}$-limited subsample (green triangles), and the S-limited samples (cyan triangles). The distribution in the COHRS sample is shown in the right panel.}
	\label{fig9} 
\end{figure*}
 

 Where $S\geq 1$, shear is strong enough to disrupt the perturbations/overdensities that seed star formation. It is in sources with $S$ above this threshold that shear is most likely to induce an overall cloud rotation. In the analysis that follows, we will consider a subset of CHIMPS sources defined by $S \geq 1$ (hence referred to as the `$S$-limited' set). This set contains the sources for which shear can potentially disrupt the gravitational collapse of clouds. We also investigate the set of sources with Galactocentric distances between 3.5 and 5.5 kpc. This '$R_\mathrm{gc}$-limited' subset comprise those sources with the greatest variation in $S$. The former subset includes 550 sources and the latter 926.

The distribution of the angles of the clouds' rotation axes with respect to the anti-parallel direction of rotation of the Galaxy ($\theta$, see Fig. \ref{skx}) is shown
in Fig.\,\ref{histo_theta}. No preferred direction of rotation is apparent in any of the CHIMPS samples considered. The \ce{^{12}CO} sources show accumulations of sources around $100^\circ$ and $300^\circ$. These directions however are not in agreement with shear-induced rotation. Fig.\,\ref{theta_galcen} show the distribution of rotation angles with the Galactocentric distance. Even in this case, no correlation is found between distance and $\theta$ (Spearman test with p-value $>> 0.001$) suggesting that even at shorter distances from the Galactic center no obvious alignment is achieved.

The solenoidal fraction, $R$, is positively correlated with the shear factor (Fig.\,\ref{sol}) in both the full CHIMPS sample (Spearman correlation coefficient $r = 0.34$, $p$-value $\ll 10^{-3}$) and the two subsamples ($R_\mathrm{gc}$-limited: $r= 0.36$; $S$-limited: $r = 0.15$, both also with p-values $\ll 10^{-3}$). This result is in agreement with the idea that stronger shear promotes the rotational modes in the gas by enhancing both the overall or partial rotation of a cloud. Both $R$ and $S$ are positively correlated with the size of the CHIMPS clouds measured by the equivalent radius (Spearman $r =0.31$   and  $0.17$, respectively, with $p$-value $\ll0.001$ for both). The shear-size correlation is much more marked in the COHRS sample, for which we find $r=0.72$. These correlations suggest that the outer, less dense layers of a molecular cloud are more sensitive to shear. Although, no correlation is found between the clouds' overall direction of rotation and the solenoidal fraction $R$ (Spearman test with p-value $>>0.001$,  Fig.\,\ref{sol}) in any of the CHIMPS samples, $R$ is positively correlated with $S$ (Spearman $r = 0.34$ with $p$-value $\ll 0.001$). This correlation is also reflected in the distribution of  $S$ with Galactocentric distance (Fig.\,\ref{S_master}) which reflects the declining values of $R$ past the edge of the Inner Galaxy shown in \cite{Rani2022}. Removing the location dependence of both $S$ and $R$ through as partial Spearman correlation test reveals positive correlation ($r=0.3$ with p-value $\ll 0.001$) between these the two quantities.




Fig. \ref{sol} highlights that, within the complex structure of CHIMPS molecular clouds, the solenoidal modes of turbulence are more likely to arise from local perturbations rather than from shear-driven global contributions.  The impact of shear on $R$ in the form of a contribution to vorticity is not detected by the magnitude of directed gradient in the distribution of velocity in first-moment maps. We also find a slightly negative correlation between the gradient magnitude and $R$ in our three samples (full CHIMPS: $r = -0.13$, $R_\mathrm{gc}$-limited: $r = -0.11$, $S$-limited $r = -0.17$, with p-values $\ll 10^{-3}$) and between the gradient magnitude and the equivalent radius of the sources (Fig. \ref{fig9}, Spearman $r = -0.15$, p-values $\ll 0.001$). The latter result indicates the higher probability of finding more well-defined gradients in smaller clouds. 


\section{Discussion and Conclusions}\label{conclusions}

Molecular clouds form as a result of the condensation of lower-density atomic ISM gas, thereby acquiring its turbulent and shear-driving motions \citep{Meidt2018, Meidt2019}. Galactic dynamics can thus either stabilise clouds \citep{Meidt2013} or compress them, facilitating star formation \citep{Jeffreson2018}. The solenoidal modes of turbulence, in particular, have been shown to be linked to reduced star formation efficiency in molecular clouds \citep{Rani2022}. The relative fraction of power in the solenoidal modes of turbulence (the solenoidal fraction, $R$) peaks at the edge of the Inner Galaxy (the region swept by the rotation of the Galactic bar) and declines with a shallow gradient with Galactocentric distance.
The strength of shear originating from the differential rotation of the Galaxy follows a similar trend (see section \ref{results}). To quantify the strength of shear, we introduced the shear factor $S$, which defines the ability of density perturbations to grow within the cloud. The higher the value of $S$, the more likely shear is to disrupt the growth of over-density and gravitational collapse \citep{Dib2012}. Both CHIMPS and COHRS clouds display the same trend with $S$ peaking at the inner edge of the Inner Galaxy and decaying with Galactocentric distance. The similarities between the distributions of $R$ and $S$ over Galactocentric distances suggest that, theoretically, shear may act as a facilitator of solenoidal turbulence. This connection is strengthened by the positive correlation between the shear parameter and the solenoidal fraction in CHIMPS clouds for which the solenoidal fraction has been calculated. The declining values of $R$ and $S$ with increased Galactocentric distances also agree with the studies of the dense gas fraction within Galactic-plane molecular clouds \citep{Longmore2013, Urquhart2013} pointing to the heightened shear of the CMZ as the cause of higher turbulent gas pressure, which raises the density threshold for star formation. A clear example of this phenomenon is the low SFE in the cloud G0.253+0.016 which appears to be caused by a prevalence of shear-driven solenoidal turbulence \citep{Federrath2016}.

The parameter $S$ exhibits overall higher values in COHRS, a fact that is compatible with the larger size of the structures traced by \ce{^{12}CO} emission ($S$ is positively correlated with the equivalent radius $R_\mathrm{eq}$) in both the CHIMPS and COHRS samples).
Despite the correlation between the shear parameter and Galactocentric distance, our study of the clouds' rotation \citep{Braine2018} in both the CHIMPS and COHRS samples found no preferred direction of rotation induced by Galactic shear (anti-parallel to the axis of rotation of the Galaxy) at all Galactocentric distances (Fig.\,\ref{theta_galcen}). The absence of correlation between rotation angles and the Galactocentric distance indicates that cloud rotation is not sensitive to the strength of shear even at distances where $S >1$ (Fig.\,\ref{S_master}). At this value of $S$, shear becomes strong enough to prevent the gravitational collapse of dense cores. Although this condition is satisfied for a large portion of CHIMPS sources and the majority of COHRS clouds at Galactocentric distances shorter than 7\,kpc, the magnitude of Galactic shear does not suffice to induce an overall cloud rotation. This finding hints at shear within the Galactic disc 
not being strong enough to induce rotation at scales at the scale of the largest clouds in the CHIMPS and COHRS samples. The rotation of these clouds thus must be imprinted at the time of their formation or subsequently. An alternative hypothesis would see most clouds at distances $< 7$\,kpc being too short-lived \citep{Jeffreson2018a, Jeffreson2018b} for shear to induce a preferred rotation in a statistically significant number of them.  

Our study also found that the magnitude of the velocity gradient in the sources is negatively correlated to both the 
$R$ and the equivalent radius. As the shear due to the Galactic rotation is not producing any preferred direction in the clouds it can not be the mechanism responsible for these correlation, and thus 'local perturbations' are the most likely candidates to explain these trends. These must be greater at lower $R_\mathrm{gc}$, however, so cloud or/and flow collisions are more influential at smaller Galactic radii.

To calculate the angle of rotation with respect to the direction of the Galactic rotation we introduce a rigid-body approximation by fitting the velocity gradient with a plane. The negative correlation between the magnitude of the velocity gradient and the size ($R_\mathrm{eq}$) of the source emphasises that larger sources may exhibit more complex structures, such as velocity flows that change significantly in different regions within the cloud.  In this framework, the velocity distribution appears to be dominated by internal gas motions originating upon cloud formation \citep{Vishniac1994}, cloud-cloud collisions \citep{Tanvir2020}  or produced by stellar feedback \citep{Fall2010}, even in the subset of clouds with shear parameters greater than unity. The slightly negative correlation between $R$ and the magnitude of the velocity gradient may imply that turbulence in smaller sources with more structured gradients is, in general, less solenoidally dominated than sources that possess more extended envelopes (both $R$ and $S$ are positively correlated with size). In the case of sources that extend over tens of parsecs, shear may increase the relative fraction of solenoidal modes in the turbulence by inducing rotation in the most extended envelopes.

Together, these results, along with the overall lack of a preferred direction of rotation in all samples we have examined would suggest that the differential rotation of the Galactic disc has little impact on molecular clouds both during their formation and after they formed.  The rotation of the clouds would thus be inherited as an imprint of the instabilities (principally Kelvin-Helmoltz instabilities) that lead to the formation of the clouds or result from their history of mergers rather than arising from instabilities induced by Galactic shear.

\section*{Acknowledgements}
G.P. was supported by the National Research Foundation of Korea through grants NRF-2020R1A6A3A01100208 \&  RS-2023-00242652.
This work was also partly supported by the Korea Astronomy and Space Science Institute grant funded by the Korea government(MSIT) (Project No. 2022-1-840-05).

\section*{Data availability}

The CHIMPS catalogue used for this paper are available from the archives of the
CHIMPS \citep{Rigby2019}. The COHRS catalogue constructed for the analysis in this article is available to download from the CANFAR archive\footnote{\url{https://www.canfar.net/citation/landing?doi=23.0019}}.



\bibliographystyle{mnras}
\bibliography{example} 

\begin{thebibliography}{}
\makeatletter
\relax
\def\mn@urlcharsother{\let\do\@makeother \do\$\do\&\do\#\do\^\do\_\do\%\do\~}
\def\mn@doi{\begingroup\mn@urlcharsother \@ifnextchar [ {\mn@doi@} {\mn@doi@[]}}
\def\mn@doi@[#1]#2{\def\@tempa{#1}\ifx\@tempa\@empty \href {http://dx.doi.org/#2} {doi:#2}\else \href {http://dx.doi.org/#2} {#1}\fi \endgroup}
\def\mn@eprint#1#2{\mn@eprint@#1:#2::\@nil}
\def\mn@eprint@arXiv#1{\href {http://arxiv.org/abs/#1} {{\tt arXiv:#1}}}
\def\mn@eprint@dblp#1{\href {http://dblp.uni-trier.de/rec/bibtex/#1.xml} {dblp:#1}}
\def\mn@eprint@#1:#2:#3:#4\@nil{\def\@tempa {#1}\def\@tempb {#2}\def\@tempc {#3}\ifx \@tempc \@empty \let \@tempc \@tempb \let \@tempb \@tempa \fi \ifx \@tempb \@empty \def\@tempb {arXiv}\fi \@ifundefined {mn@eprint@\@tempb}{\@tempb:\@tempc}{\expandafter \expandafter \csname mn@eprint@\@tempb\endcsname \expandafter{\@tempc}}}

\bibitem[\protect\citeauthoryear{Aguerre et~al.}{Aguerre et~al.}{2011}]{bolocam}
Aguerre J.~E.,  et~al., 2011, ApJS, 192, S82

\bibitem[\protect\citeauthoryear{Allen}{Allen}{1973}]{Allen1973}
Allen C.~W.,  1973, Astrophysical Quantities.
Springer

\bibitem[\protect\citeauthoryear{{Astropy Collaboration}, {Robitaille}, {Tollerud}, {Greenfield}  et~al.}{{Astropy Collaboration} et~al.}{2013}]{astropy:2013}
{Astropy Collaboration} {Robitaille} T.~P.,  {Tollerud} E.~J.,  {Greenfield}  et~al., 2013, \aap, 558, A33

\bibitem[\protect\citeauthoryear{{Astropy Collaboration}, {Price-Whelan}, {Sip{\H{o}}cz}  et~al.}{{Astropy Collaboration} et~al.}{2018}]{astropy:2018}
{Astropy Collaboration} {Price-Whelan} A.~M.,  {Sip{\H{o}}cz} B.~M.,   et~al., 2018, \aj, 156, 123

\bibitem[\protect\citeauthoryear{Bern\'{e}, Marcelino  \& Cernichora}{Bern\'{e} et~al.}{2010}]{Berne2010}
Bern\'{e} O.,  Marcelino N.,   Cernichora J.,  2010, Nature, 466, 947

\bibitem[\protect\citeauthoryear{Braine et~al.}{Braine et~al.}{2018}]{Braine2018}
Braine J.,  et~al., 2018, A\&A, 612, A1

\bibitem[\protect\citeauthoryear{Brand \& Blitz}{Brand \& Blitz}{1993}]{Brand1993}
Brand J.,  Blitz L.,  1993, A\&A, 275, 67

\bibitem[\protect\citeauthoryear{Brunt \& Federrath}{Brunt \& Federrath}{2014}]{Brunt2014}
Brunt C.~M.,  Federrath C.,  2014, MNRAS, 442, 1451

\bibitem[\protect\citeauthoryear{Brunt, Federrath  \& Price}{Brunt et~al.}{2010}]{Brunt2010}
Brunt C.~M.,  Federrath C.,   Price D.~J.,  2010, MNRAS, 403, 1507

\bibitem[\protect\citeauthoryear{Buckle, Hills, Smith  et~al.}{Buckle et~al.}{2009}]{Buckle2009}
Buckle J.~V.,  Hills R.~E.,  Smith H.,   et~al., 2009, MNRAS, 399, 1026

\bibitem[\protect\citeauthoryear{Burkert \& Bodenheimer}{Burkert \& Bodenheimer}{2000}]{Buckert2000}
Burkert A.,  Bodenheimer P.,  2000, ApJ, 543, 822

\bibitem[\protect\citeauthoryear{Churchwell et~al.}{Churchwell et~al.}{2009}]{GLIMPSE}
Churchwell E.,  et~al., 2009, PASP, 121, 213

\bibitem[\protect\citeauthoryear{Colling et~al.}{Colling et~al.}{2018}]{Colling2018}
Colling C.,  et~al., 2018, A\&A, 620, A21

\bibitem[\protect\citeauthoryear{Colombo, Rosolowsky, Ginsburg, Duarte-Cabral  \& Hughes}{Colombo et~al.}{2015}]{Colombo2015}
Colombo D.,  Rosolowsky E.,  Ginsburg A.,  Duarte-Cabral A.,   Hughes A.,  2015, MNRAS, 454, 2067

\bibitem[\protect\citeauthoryear{Colombo, Rosolowsky, Duarte-Cabral  et~al.}{Colombo et~al.}{2019}]{Colombo2018}
Colombo D.,  Rosolowsky E.,  Duarte-Cabral A.,   et~al., 2019, MNRAS, 483, 4291

\bibitem[\protect\citeauthoryear{Dame, Hartmann  \& Thaddeus}{Dame et~al.}{2001}]{Dame2001}
Dame T.~M.,  Hartmann D.,   Thaddeus 2001, ApJ, 547, 792

\bibitem[\protect\citeauthoryear{Dempsey, Thomas,   \& Currie}{Dempsey et~al.}{2013}]{Dempsey2013}
Dempsey J.~T.,  Thomas H.~S.,    Currie M.~J.,  2013, ApJS, 209, 8

\bibitem[\protect\citeauthoryear{Dib, Helou, Moore, Urquhart  \& Dariushm}{Dib et~al.}{2012}]{Dib2012}
Dib S.,  Helou G.,  Moore T. J.~T.,  Urquhart J.~S.,   Dariushm A.,  2012, ApJ, 758, 125

\bibitem[\protect\citeauthoryear{Draine}{Draine}{2011}]{Draine2011}
Draine B.~T.,  2011, Physics of the interstellar and intergalactic medium.
Princeton University Press, Oxford

\bibitem[\protect\citeauthoryear{Drazin~P.}{Drazin~P.}{1981}]{Drazin1981}
Drazin~P. R.~W.,  1981, Hydrodynamic Stability.
Cambridge Univ. Press, New York

\bibitem[\protect\citeauthoryear{Eden, Moore, Plume  et~al.}{Eden et~al.}{2017}]{Eden2017}
Eden D.~J.,  Moore T. J.~T.,  Plume R.,   et~al., 2017, MNRAS, 469, 2163

\bibitem[\protect\citeauthoryear{Elmegreen}{Elmegreen}{1993}]{Elmegreen1993}
Elmegreen B.~G.,  1993, ApJ, 411, 170

\bibitem[\protect\citeauthoryear{Elmegreen}{Elmegreen}{1995}]{Elmegreen1995}
Elmegreen B.~G.,  1995, in Yuan C.,  You Y.-H.,  eds, , Molecular Clouds Star Formation.
Singapore: World Scientific, p.~149

\bibitem[\protect\citeauthoryear{Fall, Krumholz  \& Metzner}{Fall et~al.}{2020}]{Fall2010}
Fall M.~S.,  Krumholz M.~R.,   Metzner D.,  2020, ApJ, 710, L142

\bibitem[\protect\citeauthoryear{Federrath, Rathborne, Longmore  et~al.}{Federrath et~al.}{2016}]{Federrath2016}
Federrath C.,  Rathborne J.~M.,  Longmore S.~N.,   et~al., 2016, ApJ, 832, 143

\bibitem[\protect\citeauthoryear{Fleck \& Jr}{Fleck \& Jr}{1989}]{Fleck1989}
Fleck R.~C.,  Jr 1989, AJ, 97, 783

\bibitem[\protect\citeauthoryear{Heitsch et~al.}{Heitsch et~al.}{2009}]{Heitsch2009}
Heitsch F.,  et~al., 2009, ApJ, 695, 248

\bibitem[\protect\citeauthoryear{Hunter, Sandford, Whitacker  \& Klein}{Hunter et~al.}{1986}]{Hunter1986}
Hunter J.,  Sandford M.~T.,  Whitacker R.~W.,   Klein R.~I.,  1986, ApJ, 305, 3

\bibitem[\protect\citeauthoryear{Hunter, Elmegreen  \& Baker}{Hunter et~al.}{1998}]{Hunter1998}
Hunter D.~A.,  Elmegreen B.~G.,   Baker A.~L.,  1998, ApJ, 493, 595

\bibitem[\protect\citeauthoryear{Imara \& Blitz}{Imara \& Blitz}{2011}]{Imara2011}
Imara N.,  Blitz L.,  2011, ApJ, 732, 78

\bibitem[\protect\citeauthoryear{Inoue \& Inutsuka}{Inoue \& Inutsuka}{2012}]{Inoue2012}
Inoue T.,  Inutsuka S.,  2012, ApJ, 759, 35

\bibitem[\protect\citeauthoryear{Inutsuka et~al.}{Inutsuka et~al.}{2015}]{Inutsuka2015}
Inutsuka S.,  et~al., 2015, A\& A, 580, A49

\bibitem[\protect\citeauthoryear{Jackson et~al.}{Jackson et~al.}{2006a}]{Jackson2006}
Jackson J.~M.,  et~al., 2006a, ApJS, 163, 145

\bibitem[\protect\citeauthoryear{Jackson et~al.}{Jackson et~al.}{2006b}]{GRS}
Jackson J.~M.,  et~al., 2006b, ApJS, 163, S145

\bibitem[\protect\citeauthoryear{James \& Percival}{James \& Percival}{2016}]{James2016}
James P.~A.,  Percival S.~M.,  2016, MNRAS, 457, 917

\bibitem[\protect\citeauthoryear{Jeffreson \& Kruijssen}{Jeffreson \& Kruijssen}{2018}]{Jeffreson2018}
Jeffreson S. M.~R.,  Kruijssen J. M.~D.,  2018, MNRAS, 476, 3688

\bibitem[\protect\citeauthoryear{Jeffreson et~al.}{Jeffreson et~al.}{2018a}]{Jeffreson2018b}
Jeffreson S. M.~R.,  et~al., 2018a, MNRAS, 476, 3688

\bibitem[\protect\citeauthoryear{Jeffreson et~al.}{Jeffreson et~al.}{2018b}]{Jeffreson2018a}
Jeffreson S. M.~R.,  et~al., 2018b, MNRAS, 478, 3380

\bibitem[\protect\citeauthoryear{Klein \& Woods}{Klein \& Woods}{1998}]{Klein1998}
Klein R.~I.,  Woods T.,  1998, ApJ, 497, 777

\bibitem[\protect\citeauthoryear{Kornreich \& Scalo}{Kornreich \& Scalo}{2000}]{Kornreich2000}
Kornreich P.,  Scalo J.,  2000, ApJ, 531, 366

\bibitem[\protect\citeauthoryear{Longmore, Bally  \& Testi}{Longmore et~al.}{2013}]{Longmore2013}
Longmore S.,  Bally J.,   Testi L.~o.,  2013, MNRAS, 429, 987

\bibitem[\protect\citeauthoryear{Mazumdar et~al.}{Mazumdar et~al.}{2021}]{Mazumdar2021}
Mazumdar P.,  et~al., 2021, A\& A, 650, A164

\bibitem[\protect\citeauthoryear{McKee \& Ostriker}{McKee \& Ostriker}{2007}]{McKee2007}
McKee C.~F.,  Ostriker E.~C.,  2007, ARA\&A, 45, 565

\bibitem[\protect\citeauthoryear{Meidt et~al.}{Meidt et~al.}{2013}]{Meidt2013}
Meidt S.~E.,  et~al., 2013, ApJ, 779, 45

\bibitem[\protect\citeauthoryear{Meidt et~al.}{Meidt et~al.}{2018}]{Meidt2018}
Meidt S.~E.,  et~al., 2018, ApJ, 854, 100

\bibitem[\protect\citeauthoryear{Meidt et~al.}{Meidt et~al.}{2019}]{Meidt2019}
Meidt S.~E.,  et~al., 2019, ApJ, 892, 73

\bibitem[\protect\citeauthoryear{Molinari et~al.}{Molinari et~al.}{2016}]{higal}
Molinari S.,  et~al., 2016, A\& A, 591, a149

\bibitem[\protect\citeauthoryear{Moore, Plume, Thompson  et~al.}{Moore et~al.}{2015}]{Moore2015}
Moore T. J.~T.,  Plume R.,  Thompson M.~A.,   et~al., 2015, MNRAS, 453, 4264

\bibitem[\protect\citeauthoryear{Orkisz, Pety, Gerin  et~al.}{Orkisz et~al.}{2017}]{Orkisz2017}
Orkisz J.~H.,  Pety J.,  Gerin M.,   et~al., 2017, A\&A, 599, A99

\bibitem[\protect\citeauthoryear{Park et~al.}{Park et~al.}{2022}]{Park2022}
Park G.,  et~al., 2022, ApJ, Suppl. Ser., 264, 16

\bibitem[\protect\citeauthoryear{Phillips}{Phillips}{1999}]{Phillips1999}
Phillips J.~P.,  1999, A\&AS, 134, 241

\bibitem[\protect\citeauthoryear{Pudritz \& Kevlahan}{Pudritz \& Kevlahan}{2013}]{Pudritz2013}
Pudritz R.~E.,  Kevlahan N. K.-R.,  2013, PHILOS T R SOC A, 371, 2012

\bibitem[\protect\citeauthoryear{Rani et~al.}{Rani et~al.}{2022}]{Rani2022}
Rani R.,  et~al., 2022, MNRAS, 515, 271

\bibitem[\protect\citeauthoryear{Rani et~al.}{Rani et~al.}{2023}]{Rani2023}
Rani R.,  et~al., 2023, MNRAS, 523, 1832

\bibitem[\protect\citeauthoryear{Reid, Dame, Menten  \& Brunthaler}{Reid et~al.}{2016}]{Reid2016}
Reid M.~J.,  Dame T.~M.,  Menten K.~M.,   Brunthaler A.,  2016, ApJ, 823, 77

\bibitem[\protect\citeauthoryear{Rigby, Moore, Plume  et~al.}{Rigby et~al.}{2016}]{Rigby2016}
Rigby A.~J.,  Moore T. J.~T.,  Plume R.,   et~al., 2016, MNRAS, 456, 2885

\bibitem[\protect\citeauthoryear{Rigby, Moore, Eden, Uruqhart  et~al.}{Rigby et~al.}{2019}]{Rigby2019}
Rigby A.~J.,  Moore T. J.~T.,  Eden D.~J.,  Uruqhart J.~S.,   et~al., 2019, A\&A, 632, A58

\bibitem[\protect\citeauthoryear{R\"{o}llig et~al.}{R\"{o}llig et~al.}{2011}]{Rollig2011}
R\"{o}llig M.,  et~al., 2011, EAS Publications Series, 52, 281

\bibitem[\protect\citeauthoryear{Roueff et~al.}{Roueff et~al.}{2020}]{Roueff2020}
Roueff A.,  et~al., 2020, A\&A, p.~A26

\bibitem[\protect\citeauthoryear{Sahai, Morris  \& Claussen}{Sahai et~al.}{2012}]{Sahai2012}
Sahai R.,  Morris M.~R.,   Claussen M.~J.,  2012, ApJ, 751, 69

\bibitem[\protect\citeauthoryear{Silk}{Silk}{1997}]{Silk1997}
Silk J.,  1997, ApJ, 481, 703

\bibitem[\protect\citeauthoryear{Tan}{Tan}{2000}]{Tan2000}
Tan J.~C.,  2000, ApJ, 536, 173

\bibitem[\protect\citeauthoryear{Tang et~al.}{Tang et~al.}{2013}]{Tang2013}
Tang X.~D.,  et~al., 2013, A\& A, 551, A28

\bibitem[\protect\citeauthoryear{Tanvir \& Dale}{Tanvir \& Dale}{2020}]{Tanvir2020}
Tanvir T.~S.,  Dale J.~E.,  2020, MNRAS, 494, 246

\bibitem[\protect\citeauthoryear{Toomre}{Toomre}{1964}]{Toomre1964}
Toomre A.,  1964, ApJ, 139, 1217

\bibitem[\protect\citeauthoryear{Umemoto et~al.}{Umemoto et~al.}{2017}]{FUGINI}
Umemoto T.,  et~al., 2017, PASJ, 69, 1

\bibitem[\protect\citeauthoryear{Urquhart, Moore, Schuller  et~al.}{Urquhart et~al.}{2013}]{Urquhart2013}
Urquhart J.~S.,  Moore T. J.~T.,  Schuller F.,   et~al., 2013, MNRAS, 431, 1752

\bibitem[\protect\citeauthoryear{Vishniac}{Vishniac}{1994}]{Vishniac1994}
Vishniac E.,  1994, ApJ, 428, 186

\bibitem[\protect\citeauthoryear{Watson et~al.}{Watson et~al.}{2012}]{Watson2012}
Watson L.~C.,  et~al., 2012, ApJ, 751, 123

\bibitem[\protect\citeauthoryear{Weidner, Bonnell  \& Zinnecker}{Weidner et~al.}{2010}]{Weidner2010}
Weidner C.,  Bonnell I.~A.,   Zinnecker H.,  2010, ApJ, 724, 1503

\makeatother
\end{thebibliography}




\appendix

\section{The solenoidal fraction across CHIMPS sources}\label{R_x_chimps}

To facilitate the comparison between the trend shown in Fig. \ref{S_master} and the behaviour of the solenoidal fraction, $R$ across the sample considered in this study, we reproduce the main result obtained by \cite{Rani2022}. 
As Fig. \ref{GCR} shows, $R$ peaks at the 3--4-kpc bin. Although the confirmation of this trend is required by analysing sources at lower longitudes, this result is consistent with the disc becoming stable against gravitational collapse at these radii. 

The 3--4-kpc distance bin marks the boundary between the inner Galaxy and the region of influence of the Galactic bar, which in extragalactic systems has been observed to quench star formation \citep{James2016}.

\newpage{}

\begin{figure}\label{GCR}
	\includegraphics[width=1.1\columnwidth]{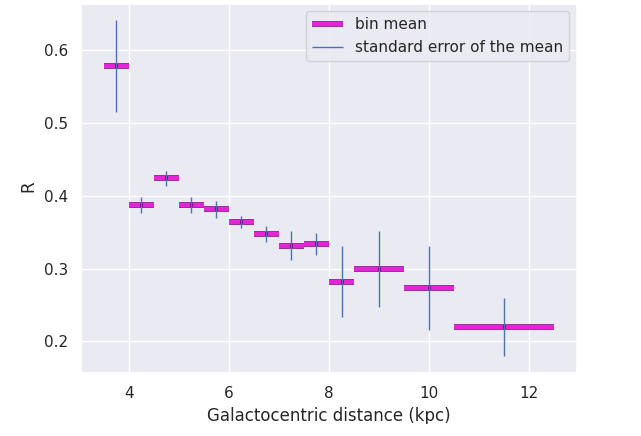}	
    \caption{
Distributions of the solenoidal fraction with Galactocentric distance in CHIMPS. The size of the bins 
is adjusted to the number of source \citep{Rani2022}. The
horizontal lines represent the mean value within each bin, 
while the vertical bars indicate the standard error of the 
mean.}
\end{figure}

%


\bsp	
\label{lastpage}
\end{document}